\documentclass[aps,prl,reprint,superscriptaddress,amsmath,amssymb,]{revtex4-1}
\usepackage{graphicx}
\usepackage{dcolumn}
\usepackage{bm}
\usepackage{color}

\usepackage{ulem}

\begin{document}


\title{Two-component Marangoni-contracted droplets: friction and shape}

\author{Adrien Benusiglio}
 \email{adrien.benusiglio@gmail.com}

\affiliation{
Department of Bioengineering, Stanford University, 450 Serra Mall, California 94305, USA}

\author{Nate Cira}
\affiliation{
Department of Bioengineering, Stanford University, 450 Serra Mall, California 94305, USA}

\author{Manu Prakash}
\affiliation{
Department of Bioengineering, Stanford University, 450 Serra Mall, California 94305, USA}

\date{\today}

\begin{abstract}
When a mixture of propylene glycol and water is deposited on a clean glass slide, it forms a droplet of a given apparent contact angle rather than spreading as one would expect on such a high-energy surface. The droplet is stabilized by a Marangoni flow due to the non-uniformity of the components' concentrations between the border and the center of the droplet, itself a result of evaporation. These self-contracting droplets have unusual properties such as absence of pinning and the ability to move under an external humidity gradient. The droplets' apparent contact angle is a function of their concentration and the external humidity. Here we study the motion of such droplets sliding down slopes, how they deform when moving at large speeds, and compare the results to normal non-volatile droplets. We precisely control the external humidity and explore the influence of the volume, viscosity, surface tension, and contact angle. We find that the droplets suffer a negligible pinning force so that for small velocities the capillary number ($\mathrm{Ca}$) is directly proportional to the Bond number ($\mathrm{Bo}$): $\mathrm{Ca}=\mathrm{Bo} \sin\alpha$ with $\alpha$ the angle of the slope. When the droplets move at larger velocities they deform when Ca exceeds a threshold, and deposit smaller droplets when $\mathrm{Ca}$ reaches twice this threshold.
\end{abstract}

\maketitle

\section*{Introduction}
When a small droplet of a pure, non-volatile liquid is deposited on a surface, it spreads until the three phase contact line around the droplet reaches an equilibrium contact angle $\theta$ with the surface. In the case that the liquid totally wets the surface (a high-energy surface for the liquid), there is no equilibrium, and the droplet continuously spreads, with a dynamically decreasing contact angle resulting from a balance of surface tension and viscous dissipation \cite{tanner1979spreading, lelah1981spreading}. If the liquid is partially wetting, the contact angle is dictated by Young's law \cite{young1805essay}. Young's law is a horizontal balance between three forces or minimization of three surface energies associated with three interfaces: the liquid/air, liquid/substrate and substrate/air interfaces \cite{deGennes-1985}. It was theoretically and experimentally shown that a non-uniform surface/air or liquid/air energy could put the droplets in motion \cite{Brochard-1989, ondarccuhu1992etalement, Chaudhury-1992}. Recently Cira \textit{et al.} \cite{Cira-2015, Cira-2014} showed that a two-component droplet of the right miscible liquids will not spread on a high-energy surface, but instead will form a well defined droplet with contact angle $\theta$. The stabilization of the droplet is due to evaporation that creates a gradient of concentration in the droplet, itself at the origin of a Marangoni flow working against the spreading force, so that the droplet is `Marangoni-contracted'. Such droplets move in response to external humidity gradients that modify their evaporation \cite{pradhan2015deposition}, and thus can attract each other \cite{Cira-2015}. This long-range attraction is made possible by the fact that surprisingly the droplets do not suffer from pinning.

The motion of a typical sessile of droplet of viscosity $\eta$, surface tension $\gamma$, volume $V$ and density $\rho$ is limited by its contact angle hysteresis $\Delta \theta$ due to microscopic geometrical or chemical inhomogeneities that induce pinning \cite{furmidge1962studies, dussan1983ability, dussan1985ability}. For example a small droplet presenting a small mean contact angle $\theta$ and contact angle hysteresis placed on an incline will only move under a gravitational force $\rho V g \sin \alpha$ larger than the pinning force $V^{1/3} \gamma \theta \Delta \theta$, with $\alpha$ the angle of the slope with the horizontal and $g$ the gravitational acceleration \cite{kawasaki1960study, dussan1983ability, quere1998drops}. Above the force threshold the droplets then move at a velocity $U$ such that the capillary number ($\mathrm{Ca}=\frac{\eta U}{\gamma}$)  is a linear function of the Bond number ($\mathrm{Bo}=\frac{V^{2/3} \rho g}{\gamma}$) times $\sin( \alpha)$ \cite{Podgorski-2001, LeGrand-2005}. For usual sessile droplets, contact angle hysteresis is strongly reduced on super-hydrophobic surfaces, and even more for liquid-infused also known as "SLIP" surfaces \cite{wong2011bioinspired, Lafuma-2011, smith2013droplet}. The droplet friction in the former case is a function of the ratio between the droplet and the oil viscosity \cite{keiser2017drop}. Here the self-contracted droplets do not suffer from pinning, as they move on a film of their own constituents. The pinning cancelation of Marangoni contracted droplets on hydrophilic substrates has hardly been reported in the literature. It was described by Huethorst and Marra \cite{huethorst1991motion}, for droplets of water in a 1-methoxy-2-propanol vapor atmosphere that maintain a constant static contact angle. In the system we present here, there is no need to place the droplets in a specific atmosphere as they self induce a Marangoni contraction through evaporation. We can control the contact angle and viscosity of the droplets by tuning their concentration and the external humidity on a large range.

In this letter we study in detail the friction of evaporative Marangoni-contracted droplets feel running down a slope, as a function of the parameters ($V$, $\theta$, $\gamma$, $\eta$), first for small slopes where droplets do not deform, then for larger slopes on which the droplet shape changes, comparing the droplets to regular non-volatile sessile droplets. First we briefly discuss how we can set $\theta$.

\section*{text}

\begin{figure}
\includegraphics[width=\columnwidth]{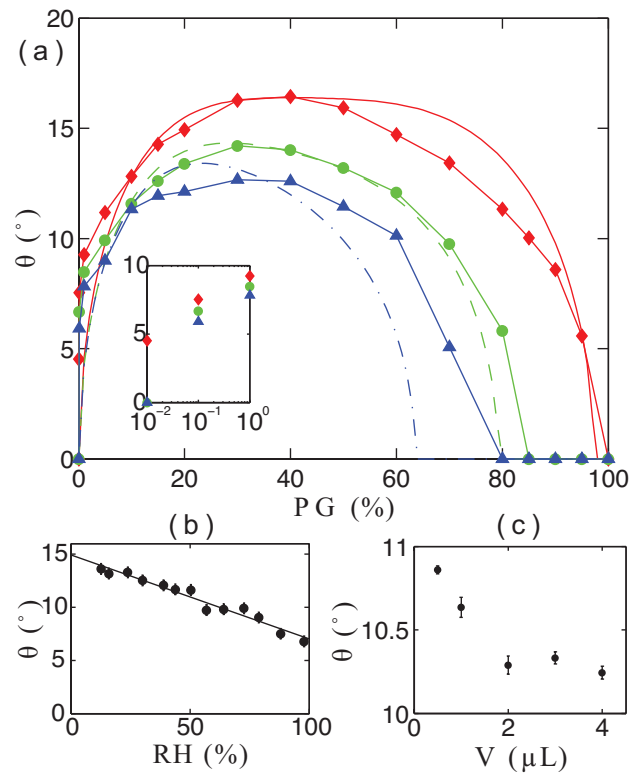}
\caption{(a) $\theta$ of 0.5 $\mu$L two-components droplets as a function of $C$ for three $RH$ (triangles: 10$\%~RH$, circles: 51$\%~RH$, diamonds: 70$\%~RH$). The lines represent the best fit of the model at each $RH$ for 10$\%~PG$. Inlet: zoom on low concentrations. (b) $\theta$ as a function of $RH$ for 10$\%~PG$, .5$\mu$L droplets. (c) $\theta$ as a function of $V$ for 10 $\% PG$ droplets at 43$\%~RH$. The error bars in b) and c) represent the standard error with minimum 3 measurements.}
\label{figure1}
\end{figure}

The droplets are composed of propylene glycol $PG$ and distilled water, with a concentration $C$ noted as $\%$ of the PG volume over the total volume. The mixture properties are extracted from the literature \footnote{The viscosity of the mixture as a function of water volume fraction $x_w$ is fitted by the polynomial: $\eta= -163 x_w^5+563 x_w^4-775.93x_w^3+551.67 x_w^2-217.87x_w +43.504$, from \cite{george2003densities}. The surface tension of the mixture as a function of $x_w$ is fitted by the polynomial: $\gamma= 25.955 x_w^5-10.211 x_w^4-5.087 x_w^3 +15.75 x_w^2+12.452 x_w^+35.47$, from \cite{karpitschka2010quantitative}.}. We first measured $\theta$ as a function of the relative humidity $RH$ and $C$.
The experiments were done in a humidity controlled chamber built in the laboratory with two sealed glove access ports. The relative humidity can be set from 10\% to 95$\%$. The droplets are deposited with a calibrated pipet on clean fully hydrophilic glass slides. The contact angle $\theta$ was measured with a reflectometry setup \cite{Allain-1985} integrated with the box, see \cite{Cira-2015} for details. Droplets of both pure liquids spread, as expected on such a high-energy surface, but adding only a small amount of PG to water (0.01 $\%$) is enough to obtain a stable $\theta$ around 5$^\circ$. The contact angle of 0.5 $\mu$L droplets as a function of $C$ for three different $RH$ is shown on Fig. \ref{figure1} (a). $\theta$ increases to a maximum and decreases back to zero as the droplet concentration is increased. The amplitude of the curve and the maximum $C$ for which a stable droplet is observed both decrease with $RH$. Varying $RH$ for $C= 10\%$ PG, we observe that $\theta$ decreases linearly with the humidity, from 14 to 6 degrees [Fig. \ref{figure1} (b)]. At a fixed humidity we observe that $\theta$ decreases slightly with $V$ from .5 to 4 $\mu$L [Fig. \ref{figure1} (c)]. In the following we will use values of $\theta$ measured for droplets of 0.5 $\mu$L. The radius of the droplet $R$ will be estimated assuming a spherical cap shape and the contact angle of $V=0.5 \mu$L droplets, as $\theta$ variation with volume only gives a 3$\%$ error on the estimation of $R$ for the larger 4 $\mu$L droplets.

A sessile droplet deposited on a surface evaporates faster from the borders than from the center because of thermal \cite{ristenpart2007influence, xu2009criterion} and mainly geometrical \cite{hu2002evaporation, eggers2010nonlocal} effects. For these two-component mixtures the droplet starts by spreading on the second time scale after deposition before building up back to a radius constant at the minute time scale. We observe under the microscope that a thin film surrounds the droplet and spreads hundreds of microns from the droplet. Interferometry measurements suggest that the film thickness increases from less than 100 nm to 300 nm as the droplet concentration is increased. As for a sessile droplet, the border of the droplet and the thin film evaporate faster than the center \cite{deegan1997capillary}. We will assume that only water is evaporating as it is 100 times more volatile than $PG$ \cite{curme1952glycols}. Due to faster evaporation and thickness, $C$ increases in the thin film and border, and remains constant in the bulk on the minute time-scale. Because $\gamma$ is monotonically decreasing with $C$, the gradient of concentration creates a gradient of surface tension that drives a Marangoni flow along the droplet surface from the border to the apex. 

\begin{figure}
\includegraphics[width=.9\columnwidth]{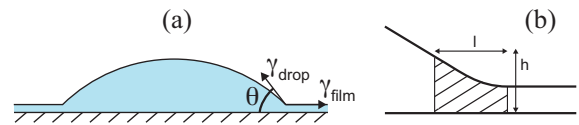}
\caption{(a) Representation of the droplet and its surrounding thin film and notations. (b) Definition of the notations describing the transition region between the bulk droplet and the thin film.}
\label{figure2}
\end{figure}

One way to model the effect of the surface tension gradient is to assume that $C$ quickly varies between the droplet and the film in a transition region of typical length $l$ and thickness $h$ (Fig. \ref{figure2}), and is quasi-constant elsewhere. The force equilibrium in the horizontal direction gives:
\begin{equation}
\gamma_\mathrm{bulk} \cos \theta = \gamma_\mathrm{film}
\end{equation}
with $\gamma_\mathrm{bulk}$ and $\gamma_\mathrm{film}$ the surface tension of the droplet and the thin film respectively. We write the conservation of water in the transition region \cite{Cira-2015}. At equilibrium, the local relative humidity above the thin film is equal to the molar concentration of water $z_\mathrm{w}$, and we assume that the evaporative flux is proportional to the difference between the local relative humidity and $RH$. There is no evaporation if $z_\mathrm{w}\leq RH$, and otherwise the evaporation happens at the rate per unit area  $\Phi_\mathrm{evap}=(z_\mathrm{w}-RH)A_\mathrm{w}$, with $A_\mathrm{w}$ the evaporative flux per unit area of pure water at 0\% $RH$. We estimate the water volume fraction in the transition region $x_\mathrm{wfilm}$ assuming that a flux per unit area $\Phi_\mathrm{in}$ of liquid of initial water volume fraction $x_\mathrm{w}$ is flowing from the droplet to the thin film. The volume of the transition region without evaporation is $V_t=\Phi_\mathrm{in} hdrdt$ with $dr$ an infinitesimal arc along the droplet perimeter and $dt$ the time to fill $V_t$. The volume of water evaporating during this time is $V_\mathrm{evap}=\Phi_\mathrm{evap}ldrdt$. The volume concentration of water in the film is then $x_\mathrm{wfilm}=\frac{x_\mathrm{w} V_t- \Phi_\mathrm{evap} l dr dt}{\Phi_\mathrm{in} h dr dt-\Phi_\mathrm{evap} l dr dt }$. Which can be rewritten:
\begin{equation}
x_\mathrm{wfilm}=\frac{x_\mathrm{w}- \left( z_\mathrm{w}-RH \right) K}{1-\left( z_\mathrm{w}-RH \right) K}
\end{equation}
with the non-dimensional parameter $K= \frac{A_\mathrm{w}l}{\Phi_\mathrm{in} h}$. For a given $RH$, choosing the best $K$ to fit the measurement, we observe that the model captures the $\theta$ trend, and predicts the maximum $RH$ up to which we observe a stable droplet [Fig. \ref{figure1} (a)]. Keeping a constant $K$ also partially predicts the amplitude variation as a function of $RH$. For droplets of 10$\%~PG$, the model predicts linear evolution of $\theta$ with $RH$, with a larger slope than what is measured. Experiments and simulations on the contact angle of Marangoni-contracted droplets  \cite{karpitschka2017marangoni}, propose a relationship $\theta \propto (RH_\mathrm{eq}-RH)^{1/3}$ with $RH_\mathrm{eq}$ the relative humidity above which the droplet spreads completely.\\

\begin{figure}
\begin{minipage}[t]{0.49\columnwidth}
\centering
 \hspace{0.2cm} (a) \\ \vspace{0.cm}
\includegraphics[width=\columnwidth]{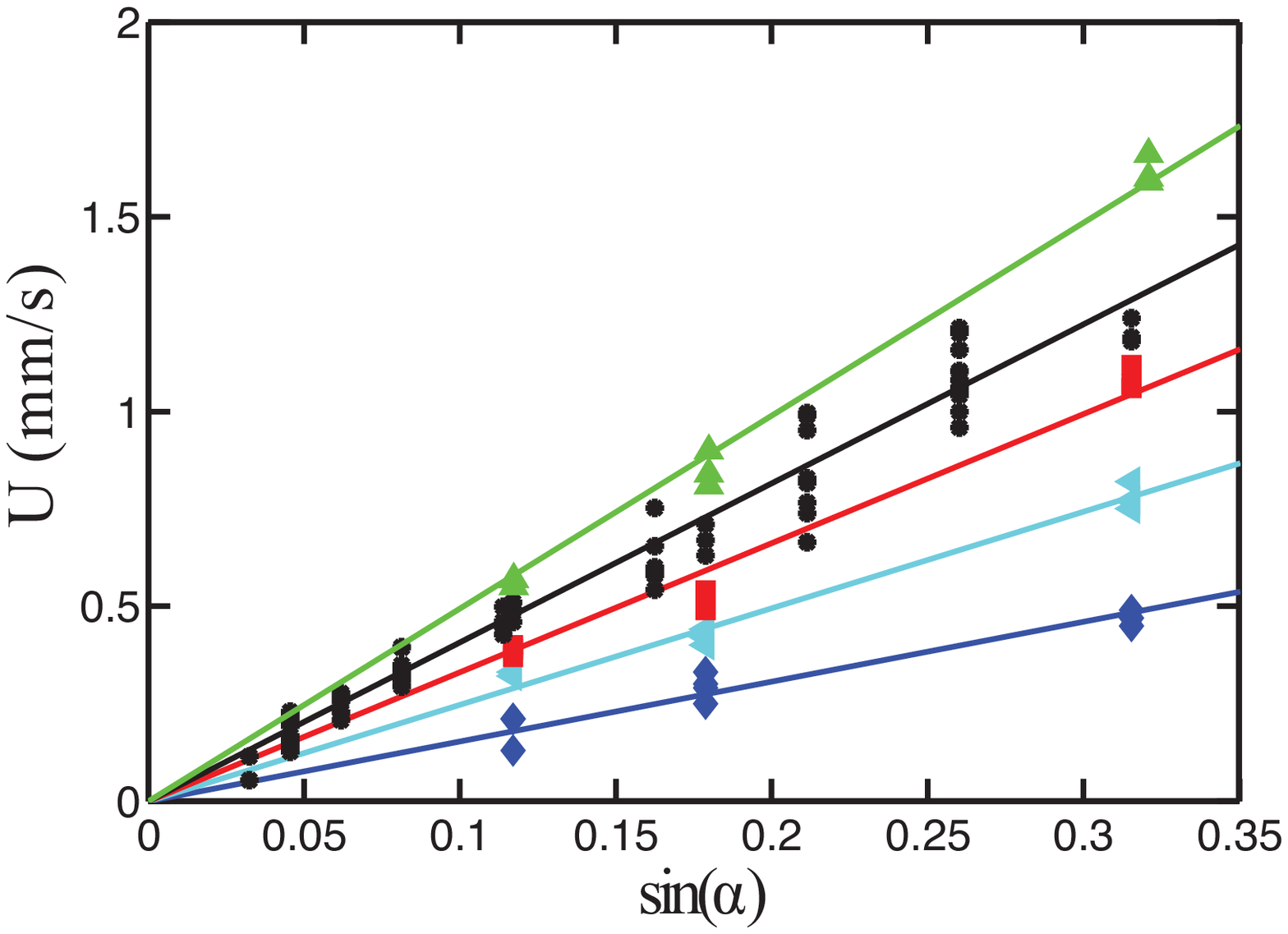}
\end{minipage}
\begin{minipage}[t]{0.49\columnwidth}
\centering
 \hspace{0.2cm} (b) \\ \vspace{0.cm}
\includegraphics[width=\columnwidth]{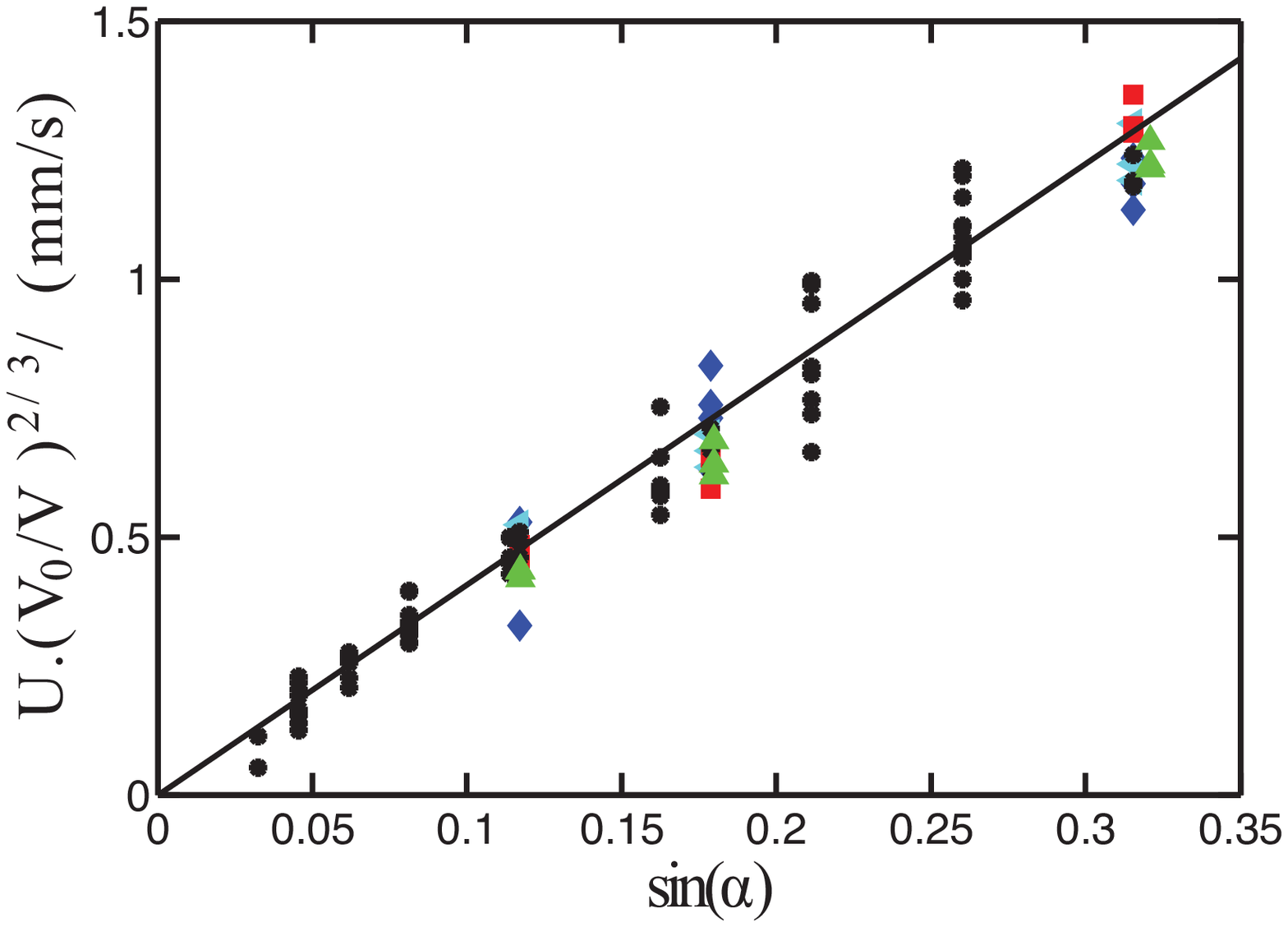}
\end{minipage}
\caption{(a) $U$ as a function of $\sin \alpha$ for $C=10\%$. The dashed lines are best linear fits, the slope increasing with the $V$= 0.25, 0.5, 0.75, 1, 1.5 $\mu$L. (b) Rescaled velocity as a function of the slope. $RH=55\%$.}
\label{figure3}
\end{figure}

Now that we either measured the parameters of the droplets as a function of $RH$ and $C$( $\theta$);,or have access to them in the literature ($\gamma$, $\eta$), we explore what is the role of these parameters on the drag of such a droplet moving down a slope. The choice of $PG$/water droplet is at first glance unpractical since when $C$ changes, $\theta$, $\gamma$ and $\eta$ change. But it in fact reveals a powerful tool to estimate the role of each parameter, as $\gamma$ is a monotonically decreasing function of $C$, when $\eta$ is monotonically increasing. The droplets are deposited on a clean glass slide placed on a slope of angle $\alpha$ from 0 to 45$^\circ$ enclosed in the humidity-controlled chamber. $V$ ranges from .25 to 10 $\mu$L. The motion is recorded from the top with a digital SLR camera. We observed that after accelerating on the millimeter scale, the droplets moved at constant $U$ in the direction of the largest slope.
For a given $V$ and $C$ we gradually increased $\alpha$ and observed that $U$ is directly linear function of $\sin \alpha$ (Fig. \ref{figure3} a). We then varied the volume of the droplets and observed that larger droplets moved faster. For a given $\alpha$ and $V$, increasing PG$\%$ we observed that $U$ monotonically decreased with the concentration, except for very small $C$ under 0.1\%.

\begin{figure}
\begin{minipage}[t]{0.65\columnwidth}
\centering
 \hspace{0.2cm} (a) \\ \vspace{0cm}
\includegraphics[width=\columnwidth]{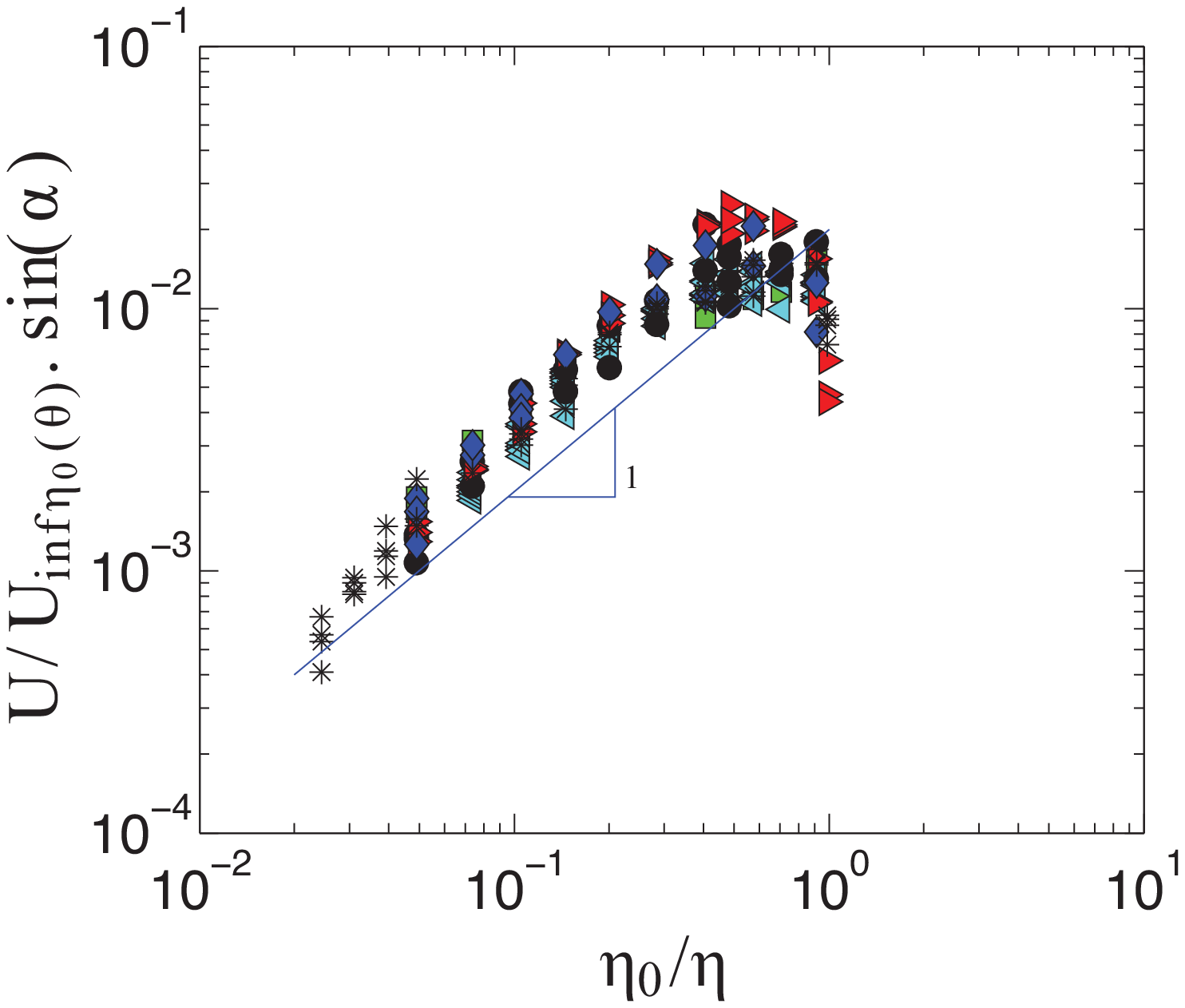}
\end{minipage}
\begin{minipage}[t]{\columnwidth}
\centering
 \hspace{0.2cm} (b) \\ \vspace{0cm}
\includegraphics[width=0.65\columnwidth]{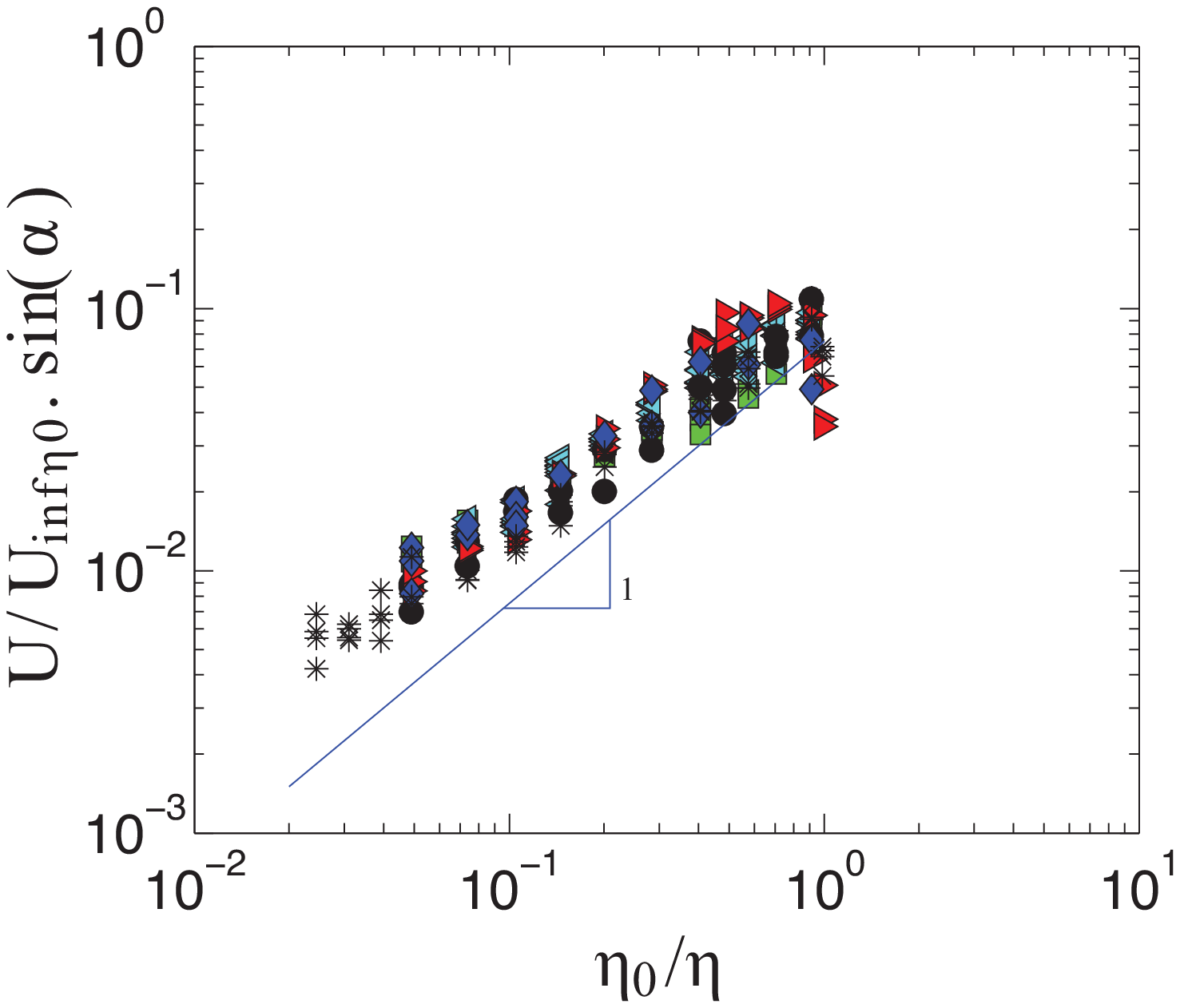}
\end{minipage}
\caption{Non-dimensional velocity as a function of the non-dimensional viscosity. (a) $A$ is a function of $\theta$, (b) $A$ is independent of $\theta$. The symbols are for different $\alpha$ an $RH$. At ambient humidity ($RH=0.50-0.55 \%$): $\circ$: $\alpha=$3.73$^\circ$; $\vartriangleright$:  19$^\circ$; $\diamond$: 7.7$^\circ$; $\square$: 13.5$^\circ$; $\vartriangleleft$: 11.3$^\circ$. At reduced humidity: $\star$: $RH=10\%$, $\alpha=$15.1$^\circ$. }
\label{figure4}
\end{figure}

Two-component droplets of typical radius $R=1$ mm move at typical velocities $U=1$ mm/s, so that the Reynolds number is equal to $\mathrm{Re}=\frac{\rho R U}{\eta}\approx 1$ (estimating Re on the thickness of the droplet $\delta \approx \theta R$, $\mathrm{Re}\approx 0.2$). When the droplets slide down the slope, they are subject to a drag force due to the gradient of velocity from the surface of the droplet to the substrate. For a sessile droplet of small contact angle and small radius compared to the capillary length $\kappa^{-1}=\sqrt{\gamma/\rho g}$, the dissipation mainly happens in the wedge close to the moving contact line. The force per unit length is written, with $U_n$ the velocity of the contact line normal to the droplet in the plane of motion: $f_\mathrm{drag}=\eta U_n \ln (b/a) / \theta$ \cite{deGennes-1985}, with $\ln (b/a)$ the logarithmic ratio of a macroscopic length scale $b$ (typically $R$) and a microscopic length scale $a$ at which the continuous matter description fails, such as the size of the molecules. A crude integration of the viscous force acting on a circular droplet is then $F_\mathrm{drag}= \pi R \eta U \ln (b/a) / \theta$. The droplet is moving due to the gravitational force projected in the direction of motion $F_\mathrm{prop}=mg \sin{\alpha}$. Equilibrating $F_\mathrm{drag}$ and $F_\mathrm{prop}$, the velocity of the droplet is $U = \frac{mg \sin{\alpha} A}{\pi \eta R}$, with $A= \theta /\ln(b/a) $. A sessile droplet presenting a contact angle hysteresis would feel an additive pinning force, and the droplet would only move above that threshold force. Here we observe that the velocity is a linear function of $\sin(\alpha)$ for droplets of 10$\%~PG$ of volumes from 0.25 to 1.5 $\mu L$ [Fig. \ref{figure3}]. The linear fit of the velocity goes to zero, showing that contrary to sessile droplets, the two-component droplets are not subject to pinning. Assuming that the droplets are spherical caps, $R \propto V^{1/3}$ and eventually $U \propto V^{2/3} \frac{\rho  g \sin{\alpha} A}{ \eta}$. We verify this scaling on Fig. \ref{figure3} (b) where we plot the velocity rescaled by $(V_0/V)^{2/3}$ with $V_0=1~ \mu$L. It indicates that the viscous dissipation happens at the border of the droplet, as with usual sessile droplets with small contact angle. From the same measurements, at 55$\%~RH$ for $C=10\%$ we extract the value of $\ln(b/a)=11.2$, while an estimation based on the size of a molecule of water $a=152$ $p$m and $R=1$ mm gives $\ln(b/a)=15.7$ and on the size of a molecule of PG $a\approx495$ $p$m gives $\ln(b/a)=14.5$.\\
We now extend the analysis to all stable (non wetting) concentrations to verify the influence of $\theta$ and $\eta$. The range of stable concentration, contact angle and viscosity is:  at $RH=55\%$ $C=1\% -80 \%$, $\theta=8-13^\circ$ and $\eta=0.9-17.04$ Cp; at $RH=10\%$ $C=0.1\%-95\%$,  $\theta=5.6-16^\circ$ and $\eta=0.84-34.17$ Cp. On Fig. \ref{figure4} (a) we rescale the data to present the non-dimensional velocity $\frac{\eta_0 R}{mg A}\frac{U}{\sin \alpha}=\frac{U}{U_{\mathrm{inf}\eta_0} \sin \alpha} \propto \eta_0/\eta$, with $\eta_0$ the viscosity of pure water and $U_\mathrm{inf} \eta_0$ the theoretical velocity of non-deformed drops on a vertical surface, if they had the viscosity of pure water, as a function of $\eta_0 / \eta$. We observe that the rescaled velocity is not a monotonically increasing function of the rescaled viscosity, as we would expect, especially for small $\eta$. This could come from the fact that the volume in which the dissipation happens is not a function of $\theta$, if the transition region between the droplet and the thin film is not affected by the apparent $\theta$. We test this hypothesis, non-dimentionalizing the velocity with a fixed value of $A$, that is not a function of the apparent $\theta$, and plot the non-dimensional velocity as a function of the non-dimensional viscosity in Fig. \ref{figure4} (b). We now observe that the rescaled velocity is linear function of $\eta$ on more than one decade, which shows that the contact angle of the droplets has no influence on the drag other than the influence it has on $R$. We also observe a slight deviation from linearity for the largest $\eta$, suggesting a reduced drag at large slope not captured by the model. It could be induced by a thicker thin film for large viscosities.\\

\begin{figure}
\includegraphics[width=\columnwidth]{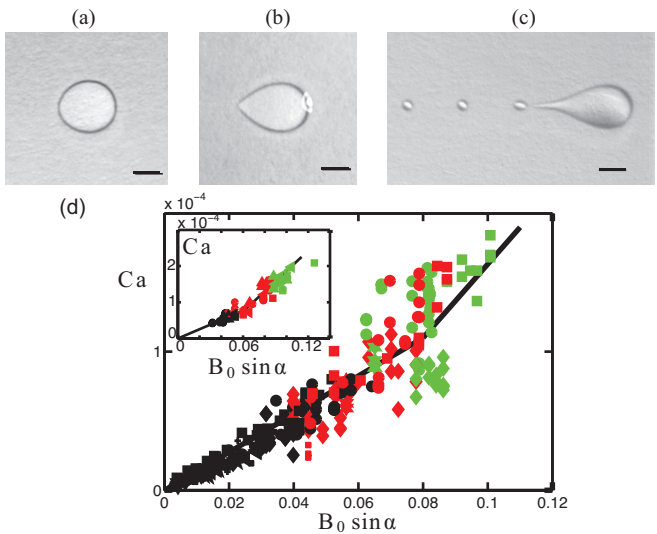}
\caption{Shapes of a 1$\mu L$, $C=30\%$ droplet as the slope is increased: (a) oval, (b) cornered, (c) pearling (the solid bar represents 2 mm). (d) $Ca$ as a function of $Bo\sin \alpha$. Black, red and green symbols represent respectively rounded, cornered and pearling droplets.The symbols represent different concentrations and volumes: the volumes are only indicated for the PG 10 droplet represented by triangles ( $\triangledown$: 0.25 $\mu$L, $\triangle$: 0.5 $\mu$L, $\vartriangleright$: 0.75 $\mu$L, $\vartriangleleft$: 1 $\mu$L, $\diamond$: 1.5 $\mu$L, hexagram: 2 $\mu$L). $\square$: $C=40\%$ 1.5 $\mu$L; pentagram: $C=5\%$, 1.5 $\mu$L. $+$: $C=1\%$; $\circ$: $C=20\%$, various volumes between 0.5 and 2 $\mu$L. Insert: $C=40\%$ droplets only, the symbols represent different volumes: $\circ$: 1.5 $\mu$L,  $\triangle$: 3 $\mu$L, $\vartriangleleft$: 5 $\mu$L, $\square$: 10 $\mu$L.}
\label{fig:deformation}
\end{figure}

When a sessile droplet is running down a slope at larger and larger velocity, its back deforms from an oval shape to a corner shape, to a cusp emitting smaller droplets \cite{Podgorski-2001, LeGrand-2005}. The two-component droplet is stabilized by an internal flow that may reduce the deformation. We now study the motion and shape of the two-component droplets on large slopes. For any $C$, as we increase $\alpha$, the back of the droplet deformed from an oval [Fig. \ref{fig:deformation} (a)] to a cornered (b), to a cusped droplet emitting smaller droplets (c). The successive shapes were similar to what is observed for a sessile droplet, with the notable exception that the droplet do not present parallel sides between the advancing and receding perimeters, showing again that the droplets do not present contact angle hysteresis \cite{dussan1983ability, huethorst1991motion}. We discuss a simplified model to capture the essence of the equilibrium leading to deformation. Along the virtual contact line of the droplet, on the receding side the drag force per unit length is balanced by surface tension such as $\eta \ln (b/a) U \cos{\Phi}= \gamma$, with $\Phi$ the angle between the normal to the droplet and $U$ in the plane of motion. If $\eta \ln (b/a) U < \gamma$ the back of the droplet should stay non-deformed, and otherwise  $\cos{\phi}= \frac{\eta U}{\gamma}= 1/Ca$, which gives us the threshold for deformation  $\mathrm{Ca}>1$. Plotting Ca versus $\mathrm{Bo} \sin{\alpha}$ on Fig. \ref{fig:deformation} (d) we observe that all the data collapses on a directing curve, linear for small bond numbers as discussed in the introduction, and that goes to zero for small velocities because there is no contact angle hysteresis (here we define $Bo$ as $Bo=\frac{V \rho g}{R \gamma}$ because $\theta$ and thus $R$ are function of the concentration). As $\sin{\alpha}$ was increased, the droplets started to deform when $\mathrm{Ca}>0.5 \times 10^{-4}$ and even emit smaller droplets when $\mathrm{Ca}>1 \times 10^{-4}$, and the slope of the curve increased, similarly to what is observed for sessile droplets \cite{Podgorski-2001, LeGrand-2005}. The value of $Ca$ for which deformation becomes noticeable is constant in a large range of volumes [Fig. \ref{fig:deformation} (d) insert], but seems to be weakly dependent on $C$. This observation is consistent with the fact that the drag force a droplet feels is slightly smaller than expected as $\eta$ increases, as seen on Fig. \ref{figure4}.


\section*{Conclusion}
Two-component mixtures of well chosen miscible liquids do not spread on high-energy surfaces but rather form stable droplets. Non-uniform concentration due to evaporation creates a Marangoni flow that stabilizes the droplets. A simple force balance model coupled with an estimation of the gradient of concentration gives a good picture of the Marangoni contraction process. Like typical sessile droplets, the two-component droplet velocity down a slope is a linear function of $\mathrm{Bo} \sin{\alpha}$, but contrary to sessile droplets, they do not present contact angle hysteresis, so that $\mathrm{Ca}=\mathrm{Bo} \sin{\alpha}$. Contrary to sessile droplets, the dissipation does not seem to be influenced by the static apparent contact angle of the droplets in the range of this study. The absence of pinning is surprising, and the microscopic process making it possible remain to be determined but the presence of a thin film around the droplet seems to play a role. Like with sessile droplets, when these droplets run down slopes of increasing angle, their receding perimeter deforms from a circular arc to a cone at a given Ca and later deforms into a cusp that deposit smaller droplets, at twice the value of Ca.


\begin{acknowledgments}
  \end{acknowledgments}

\bibliography{bib_DropDrag}
\end{document}